\documentclass[aps,prl,twocolumn,superscriptaddress,floatfix,amsmath,amssymb,nofootinbib,longbibliography]{revtex4-2}

\usepackage[T1]{fontenc}
\usepackage[utf8]{inputenc}
\usepackage[english]{babel}
\usepackage{amssymb,amsthm,amsmath,amstext,amsbsy,amsopn}
\usepackage{bbm}
\usepackage{graphicx}
\usepackage{isotope}
\usepackage{float}
\usepackage[dvipsnames]{xcolor}
\usepackage{orcidlink}
\usepackage{hyperref}
\hypersetup{
    colorlinks = true,
    citecolor  = blue,
    linkcolor  = blue,
    urlcolor  = blue
}

\newcommand{\la}{\langle}
\newcommand{\ra}{\rangle}
\newcommand{\ket}[1]{|#1\rangle}
\newcommand{\bra}[1]{\langle#1|}
\newcommand{\Xmax}[1]{\ensuremath{#1_\text{max}}}

\newcommand{\cluster}[1]{\ensuremath{\mathcal{T}_{#1}}}

\newcommand{\elem}[2]{{$^{#2}$}\text{#1}}

\newcommand{\ai}{\textit{ab initio}}
\newcommand{\ie}{\textit{i.e.}}
\newcommand{\eg}{\textit{e.g.}}

\newcommand{\magicint}{1.8/2.0 (EM)}
\newcommand{\deltago}{$\Delta$-N$^2$LO$_\text{GO}$}

\newcommand{\MeV}{\text{MeV}}

\newcommand{\bigO}{O}

\begin{document}

\allowdisplaybreaks

\title{
High-precision ab initio calculations of nuclear binding energies: \\ Tin isotopes from dripline to dripline
}

\author{U.~Vernik \orcidlink{0000-0002-3717-945X}}
\email{urban.vernik@tu-darmstadt.de}
\affiliation{Technische Universit\"at Darmstadt, Department of Physics, 64289 Darmstadt, Germany}
\affiliation{ExtreMe Matter Institute EMMI, GSI Helmholtzzentrum f\"ur Schwerionenforschung GmbH, 64291 Darmstadt, Germany}
\affiliation{KU Leuven, Instituut voor Kern- en Stralingsfysica, 3001 Leuven, Belgium}

\author{P.~Demol \orcidlink{0000-0003-2511-7179}}
\email{pepijn.demol@ulb.be}
\affiliation{KU Leuven, Instituut voor Kern- en Stralingsfysica, 3001 Leuven, Belgium}
\affiliation{%
Institut d’Astronomie et d’Astrophysique, Université Libre de Bruxelles, 1050 Brussels, Belgium
}
\affiliation{%
Brussels Laboratory of the Universe -- BLU-ULB, 1050 Brussels, Belgium}

\author{T.~Duguet \orcidlink{0000-0002-7596-3851}}
\email{thomas.duguet@cea.fr} 
\affiliation{KU Leuven, Instituut voor Kern- en Stralingsfysica, 3001 Leuven, Belgium}
\affiliation{IRFU, CEA, Universit\'e Paris-Saclay, 91191 Gif-sur-Yvette, France}

\author{A.~Tichai \orcidlink{0000-0002-0618-0685}}
\email{alexander.tichai@tu-darmstadt.de}
\affiliation{Technische Universit\"at Darmstadt, Department of Physics, 64289 Darmstadt, Germany}
\affiliation{ExtreMe Matter Institute EMMI, GSI Helmholtzzentrum f\"ur Schwerionenforschung GmbH, 64291 Darmstadt, Germany}
\affiliation{Max-Planck-Institut f\"ur Kernphysik, Saupfercheckweg 1, 69117 Heidelberg, Germany}

\begin{abstract}
The location of the neutron drip line in tin isotopes has important consequences for our fundamental understanding of nuclear structure and nuclear forces as well as for astrophysical nucleosynthesis. Performing high-precision \ai{} calculations of even-even tin isotopes from $N=50$ to $N=126$ based on chiral two- and three-nucleon interactions, the predicted drip-line location is found to be highly sensitive to the employed nuclear interactions and to exhibit tension with recent energy-density-functional predictions. On the neutron-deficient side, results are consistent with extrapolated two-neutron separation energies constrained by recent Penning-trap mass measurements.
\end{abstract}

\maketitle
\paragraph{Introduction.--}
First-principles calculations of atomic nuclei are rapidly progressing towards heavy and exotic nuclei~\cite{Hebe203NF,Herg20review}.
Combining nuclear two- and three-body interactions derived from chiral effective field theory ($\chi$EFT)~\cite{Epel09RMP,Mach11PR,Hamm20RMP} with systematically improvable many-body expansion methods allows a controlled microscopic description of nuclear observables with predictive power and rigorous quantification of theoretical uncertainties. This has enabled extensive applications throughout the lower end of the nuclear chart, targeting a variety of nuclear properties ranging from binding energies~\cite{Herg13MR,Stroberg2021,Hu2025texas}, nuclear radii~\cite{Novario2020a,Mueller2025,Wolfgruber2025,Gustafsson2025,Demol2026bcc}, low-lying spectroscopy, all the way to electromagnetic and weak responses~\cite{Mior16dipole,Bonaiti2024,Porro2025,Zhou2025, Li2026Beta} well as the description of giant resonances~\cite{Porro2024response1,Porro2024response2,Porro2024response3,Porro2024response4,Porro2025}.
With increasing computing power, \ai{} calculations have started to target nuclei with mass number $A \gtrsim 100$~\cite{Morr17Tin,Arthuis2020a,Tichai2024bcc,Arthuis24a,Door2025ytterbium,hild25NLEFTSn,Demol2026bcc}, with flagship applications to doubly-magic \elem{Pb}{208} and \elem{Pb}{266}~\cite{Hu2021lead,Bonaiti2025heavy} nuclei.

Microscopic calculations in the mid-mass regime are commonly achieved through the use of many-body expansion methods, in which dominant particle-hole correlations are added on top of a many-body reference state capturing bulk properties of the target nucleus~\cite{Herg20review,Soma11GGFform,Tich16HFMBPT,Tichai2020review,Hage14RPP,Tsuk11IMSRG,Tsuk12SM}.
Based on their mild computational scaling, a variety of non-perturbative many-body frameworks such as self-consistent Green's function~\cite{Dickhoff:2004xx,Soma:2020xhv,Soma:2020dyc}, coupled-cluster (CC) theory~\cite{Hage14RPP,Hage16Ni78} and the in-medium similarity renormalization group (IMSRG)~\cite{Tsuk11IMSRG,Bogn14SM,Herg14MR,Herg17PS} offer a path towards heavy nuclei.
In their basic formulation, these approaches are limited to closed-shell nuclei such that their application away from shell closures requires further methodological developments.
The most widely used technique is the valence-space formulation of the IMSRG (VS-IMSRG), where an active-space Hamiltonian is constructed and used in a large-scale diagonalization~\cite{Bogn14SM,Stroberg2019,Stroberg2021}.
Alternatively, using symmetry-breaking reference states offers to keep a moderate computational scaling at the price of generalizing the many-body algebra at play.
This rationale has been successfully applied by breaking particle number and/or angular momentum to account for static superfluid and/or quadrupolar correlations that govern strongly correlated nuclei~\cite{Soma11GGFform,Soma14GGF2N3N,Tsuk11IMSRG,Sign14BogCC,Duguet2014su2,Duguet2015u1,Tichai18BMBPT,Yao2020,Novario2020a,Hagen2022PCC}.

While remaining tractable, the computational demand increases significantly when breaking symmetries and precision calculations in heavy open-shell nuclei constitute a severe challenge.
The recently proposed Bogoliubov coupled-cluster (BCC) framework~\cite{Sign14BogCC} is presently advanced by including, for the first time, leading triples corrections in the calculation of nuclear binding energies. Compared to previous results~\cite{Tichai2024bcc,Marino2026,Demol2026bcc}, the many-body uncertainty is reduced by a factor of ten, leading to a residual error of less than $1 \%$. With this, high-precision calculations based on modern chiral EFT interactions are performed along the complete tin ($Z=50$) isotopic chain (\elem{Sn}{96-180}) from the neutron-deficient $N=50$ shell closure all the way to the predicted neutron dripline.
Results reveal the sensitivity of the neutron dripline location to interactions details, as well as an emerging conflict between \ai{} and energy density functional (EDF) predictions.
This discrepancy is important and will have to be resolved given the large impact neutron-rich tin isotopes have on nuclear $r$-process nucleosynthesis networks calculations and the resulting abundance patterns~\cite{Storbacke2024, Kuske2025}.

\paragraph{Many-body framework.--}
\label{s:manybody}
The aim of {\it ab initio} nuclear structure calculations is to find the solution of the many-body Schrödinger equation $H\ket{\Psi_n} = E_n\ket{\Psi_n} \,$ based on a realistic $\chi$EFT nuclear Hamiltonian.
In this work, the Bogoliubov CC framework is employed. It relies on a particle-number breaking Bogoliubov vacuum $|\Phi\rangle = \mathcal{C}\prod_k\beta_k|0\rangle$ as a reference state that is built from a set of quasi-particle operators $\{\beta_p,\beta^\dagger_p\}$. The latter is linked to a set of single-particle operators $\lbrace c_p,c^\dagger_p\rbrace$ through a unitary Bogoliubov transformation (see Ref.~\cite{RingSchuck}): $\beta_k \equiv \sum_p U_{pk}^* c_p +  V_{pk}^* c_p^{\dagger}$.
The transformation matrices $(U,V)$ are obtained from a variational solution of the Hartree-Fock-Bogoliubov (HFB) mean-field equations that deliver a set of Bogoliubov quasi-particle energies $\{E_k\}$ at the same time. As the HFB reference state breaks particle-number conservation, the Hamiltonian is replaced by the grand-canonical potential $\Omega \equiv H - \lambda_N N - \lambda_Z Z\,$ where the neutron and proton chemical potentials $\lambda_N$ and $\lambda_Z$ serve as Lagrange multipliers to constrain the average number of neutrons and protons to match the physical values of the target system. Reference elementary excitations of the system are obtained as quasi-particle excitations of the reference state $|\Phi^{\alpha\beta\dots}\rangle \equiv \beta^{\dagger}_{\alpha}\beta^{\dagger}_{\beta}\dots|\Phi\rangle$. 

In BCC theory, the fully correlated ground state is obtained via the action of an exponential wave operator $|\Psi\rangle \equiv e^\mathcal{T}|\Phi\rangle$ onto the reference state,  where the connected cluster operator $\mathcal{T} \equiv \sum_{n} \mathcal{T}_n$ involves components of different excitation rank $n$~\cite{Sign14BogCC}
\begin{equation}\label{e:Clusteramp}
    \mathcal{T}_n \equiv \frac{1}{(2n)!}\sum_{k_i\dots k_{2n}}t_{k_1\dots k_{2n}}\beta^\dagger_{k_1}\dots\beta^\dagger_{k_{2n}}\, .
\end{equation}
The cluster amplitudes $t_{k_1\dots k_{2n}}$ are the unknowns of the problem and must be solved for numerically by left-decoupling the \emph{similarity-transformed grand potential}
$\bar{\Omega}_N \equiv e^{-\mathcal{T}}\Omega_N e^{\mathcal{T}}$, with $\Omega_N \equiv \Omega - \langle \Phi | \Omega | \Phi \rangle$,
from elementary excitations, \ie{},
$\la \Phi^{\alpha\beta\dots}| \bar \Omega_N | \Phi \ra = 0$.
The decoupling is achieved through an iterative procedure~\cite{Hage14RPP}.
Given the cluster amplitudes, the correlation energy is obtained as $E_\text{corr.} =\bra{\Phi}{\bar \Omega}_N\ket{\Phi}_{\text{c}}$.

In previous works~\cite{Tichai2024bcc,Demol2026bcc}, the BCC amplitude equations were solved within the singles and doubles (BCCSD) approximation where $\mathcal{T} \approx \mathcal{T}_1 + \mathcal{T}_2$. At a mild computational cost of $\bigO(N^6)$ (where $N$ denotes the system size), BCCSD resums all correlations up to third-order in perturbation theory plus a subset of contributions up to infinite order. This can be shown empirically to account for about $90$\% of the correlation energy~\cite{Hage14RPP}.
To improve the accuracy of the BCCSD approximation, leading contributions from $\cluster{3}$ amplitudes~\cite{Bind13expl3NLCCSD(T),Miorelli2018,Novario2020a} are presently included in Bogoliubov CC calculations for the first time.
As the full inclusion of the $\cluster{3}$ operator is computationally prohibitive, the simpler non-iterative BCCSD[T] approximation is employed that is based on a set of converged $\cluster{2}$ amplitudes (see End matter (EM) for details):
starting from a set of converged $\cluster{2}$ amplitudes, the $\cluster{3}$ amplitudes are approximated as
\begin{equation}
    t_{\alpha\beta\gamma\delta\epsilon\zeta} \equiv -\mathcal{A} \sum_{k_{1}} \frac{\Omega^{31}_{\delta\epsilon\zeta k_{1}}\,t_{\alpha\beta\gamma k_{1}}}{\Delta E_{\alpha\beta\gamma\delta\epsilon\zeta}}
    \label{e:triplesBCCSDT1a}\, ,
\end{equation}
where $\mathcal{A}$ denotes the anti-symmetrizer ensuring fermionic symmetry, $\Omega^{31}$ a specific component of $\Omega$ expressed in the quasi-particle basis~\cite{Sign14BogCC} and where the energy denominator involves the sum of quasi-particle energies $\Delta E_{\alpha\beta ...}\equiv E_\alpha + E_\beta + ...$. The corresponding BCCSD[T] energy correction reads as
\begin{align}
    E_\text{t} 
    &= -\frac{1}{6!}\sum_{\alpha\beta\gamma\delta\epsilon\zeta}\,t^{*}_{\alpha\beta\gamma\delta\epsilon\zeta}\,\Delta E_{\alpha\beta\gamma\delta\epsilon\zeta}\,t_{\alpha\beta\gamma\delta\epsilon\zeta} \, .
    \label{eq:e_t{T}}
\end{align}
As detailed in the EM, the computationally most expensive part is the formation of the triples amplitudes themselves, involving $\bigO(N^7)$ operations and hence exceeds the cost of solving the BCCSD amplitude equations. The proposed BCCSD[T] framework does not require the storage of multiple copies of the $\cluster{3}$ amplitudes, such that the memory consumption is still driven by the \cluster{2} amplitudes and the storage of the corresponding DIIS history~\cite{Pulay1980diis}. A detailed diagrammatic derivation of the full BCCSDT equations is given in Refs.~\cite{Demol2024Thesis, Vernik2024Master}.

\begin{figure*}[t!]
    \centering
    \includegraphics[width=0.9\textwidth]{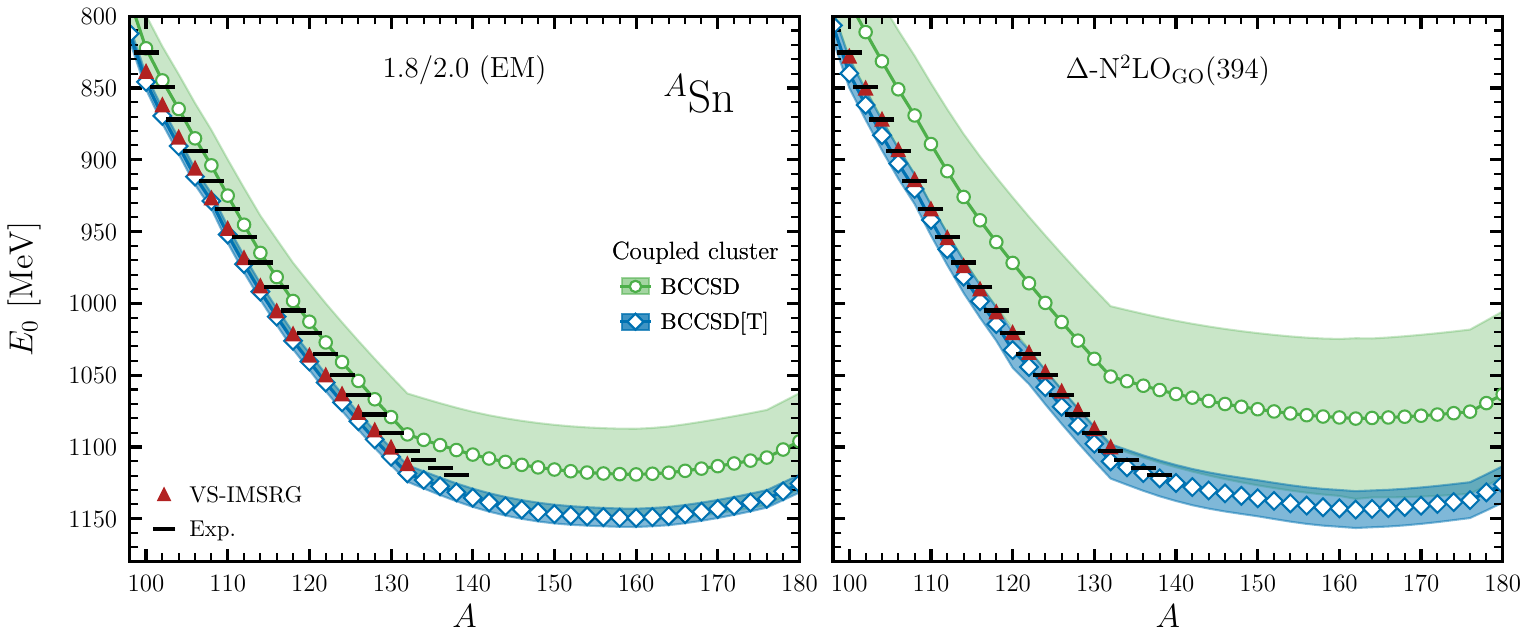}
    \caption{Ground-state energies of even-even tin isotopes with $A=100-180$ using BCCSD and BCCSD[T] calculations based on the \magicint{} (left) and \deltago{} (right) $\chi$EFT Hamiltonians. Results from VS-IMSRG calculations are taken from Ref.~\cite{Gustafsson2025}. See the main text for the discussion of many-body uncertainties.}
    \label{fig:tin_gse}
\end{figure*}

\paragraph{High-precision calculations along the tin chain.--}

Two different $\chi$EFT Hamiltonians are presently used: the \magicint{} interaction from Ref.~\cite{Hebe11fits} and the $\Delta$-full interaction \deltago{} from Ref.~\cite{Jian20N2LOGO} with a cutoff of $\Lambda = 394 \, \MeV$.
Many-body operators are represented using a one-body spherical harmonic oscillator (HO) basis characterized by the frequency $\hbar \omega=12 \, \MeV$ and truncated to include 15 major shells, \ie{}, $\Xmax{e} = (2n + l)_\text{max}=14$.  Three-body matrix elements are limited to three-body basis states characterized by $E^{(3)}_\text{max} = e_1 + e_2 + e_3 \leq 24$, which is sufficient to reach convergence of many-body observables in the target mass regime~\cite{Miyagi2021,Tichai2024bcc}. Three-nucleon interactions are approximated using normal-ordering techniques to build an effective (nucleus-dependent) two-nucleon interaction from the reference state. This was shown to introduce only moderate errors of the order of $1-2\%$ on bulk properties~\cite{Hage07CC3N,Roth12NCSMCC3N,Bind13expl3NLCCSD(T),Ripoche2020,Frosini2021}.
The residual BCCSD[T] uncertainty is estimated to be $1\%$ of the BCCSD[T] correlation energy. Compared to the common $10\%$ uncertainty at BCCSD level, this reduces the uncertainty from about $30\, \MeV$ to less than $3\, \MeV$ in \elem{Sn}{100}.
This is comparable to state-of-the-art IMSRG calculations with an approximate account of three-body operators, see Refs.~\cite{Heinz2020,Heinz2025}.
In addition, the residual one-body basis uncertainty is estimated to be of the order of $3\, \MeV$ based on explicit variations of the underlying HO frequency, and $\Xmax{e}$ value.

Figure~\ref{fig:tin_gse} compares BCCSD and  BCCSD[T] ground-state energies to experimental values for all even-mass \elem{Sn}{100-180} isotopes using the \magicint{}  (left) and \deltago{} (right) Hamiltonians. Results from VS-IMSRG(2) calculations in the same model space and a slightly different HO frequency $\hbar \omega = 16 \, \MeV$~\cite{Gustafsson2025} are also displayed.
For both Hamiltonians, the attractive contribution from triples corrections is significant and brings total ground-state energies systematically much closer to VS-IMSRG(2) values that are presently taken as a reference. The size of the triples corrections ranges between 30 and 50 \MeV{} for the \magicint{} Hamiltonian. This is consistent with the expected $10 \%$ uncertainty empirically found at CCSD level~\cite{Hage14RPP}.
While the relative size of the triples compared to BCCSD is similar for the \deltago{} interaction, their magnitude is enlarged by a factor of two owing to the higher built-in resolution scale. The larger BCCSD[T] correlation energies are compensated for by a smaller HFB reference-state energy compared to the \magicint{} Hamiltonian. Total BCCSD[T] energies from both Hamiltonians agree with each other within their respective many-body uncertainties.

The inclusion of triples corrections reduces the many-body uncertainty significantly throughout the isotopic chain, improving the precision of the overall calculation by about a factor of ten. The residual uncertainty at BCCSD[T] level is due the effects of non-perturbative triples, missing higher-body correlations ($\cluster{k \geq4}$), one-body basis incompleteness and the lack of particle-number symmetry restoration in open-shell nuclei~\cite{Tichai2024bcc}.
The total many-body uncertainty is about $5\, \MeV$ for \magicint{} and $10 \, \MeV$ for the \deltago{}. This corresponds to an approximation error of less than $1 \, \%$ on ground-state energies up to mass numbers $A=180$, thus allowing for precision tests of chiral Hamiltonians in exotic heavy-mass nuclei.

As seen on Fig~\ref{fig:tin_gse}, BCCSD[T] ground-state energies overpredict VS-IMSRG(2) values, which is again consistent with the general findings that the IMSRG(2) truncation resides in between CCSD and CCSD$+$T$_X$---independently of the particular triples approximation employed. The inclusion of triples effects in the IMSRG revealed an attractive effect of about $2 \%$ of the correlation energy in medium-mass calcium isotopes, see Ref.~\cite{Heinz2025}, suggesting that the gap between BCCSD[T] and VS-IMSRG(2) is reduced once three-body operators are taken into account in the latter~\cite{Heinz2020}.
 Bogoliubov CC results highlight that the inclusion of triples corrections does not necessarily improve the reproduction of experimental data as indicated by their root mean square (RMS) error,
$\sigma_\text{RMS} \equiv \sqrt{\frac{1}{N} \sum_{i}(E_i^{\text{th.}} - E_i^\text{exp.})^2}$ due to the uncertainty on the nuclear interaction itself.
In fact, BCCSD ground-state energies obtained from the \magicint{} Hamiltonian are in better agreement with measurements for most nuclei ($\sigma_\text{RMS}= 10.1 \, \MeV$) than BCCSD[T] ones ($\sigma_\text{RMS} = 17.9 \, \MeV$). This holds particularly, for neutron-deficient isotopes near \elem{Sn}{102} where both BCCSD[T] and VS-IMSRG(2) systematically overbind experimental data. Moving towards the $N=82$ shell closure, BCCSD and BCCSD[T] values have comparable deviation from experiment. Eventually, the \magicint{} Hamiltonian overpredicts experimental ground-state energies by about $2-3 \%$ in neutron-rich tin isotopes up to the last available data point in \elem{Sn}{138}. For the \deltago{} Hamiltonian characterized by a higher intrinsic resolution scale, the RMS error at BCCSD level is large ($\sigma_\text{RMS} = 48.3 \, \MeV$) and the inclusion of triples correlations improves the situation drastically leading to essentially converged ground-state energies in very good agreement with experimental masses ($\sigma_\text{RMS} = 8.9 \, \MeV$). 

\paragraph*{Nuclear structure at the $N=50$ shell closure.--}
\label{s:results}

Having established the BCCSD[T] approximation, the $N=50$ shell closure in \elem{Sn}{100} is analyzed in detail by investigating two-neutron separation energies 
\begin{align}
    S_{2n}(N,Z) \equiv E(N,Z) - E(N-2,Z) \, ,
\end{align}
along a set of even-even neutron-deficient tin isotopes~\cite{Mougeot2021}.
The latest Atomic Mass Evaluation from 2020 (AME 2020) includes measurements down to \elem{Sn}{103}, whereas binding energies and two-neutron separation energies of lighter systems employ mass extrapolations down to \elem{Sn}{100}.
Based on recent high-precision mass measurement using the LEBIT Penning trap, the experimental mass of \elem{Sn}{101,103} was extracted at much improved precision, thus enabling a more robust (Bayesian) extrapolation towards even more neutron-deficient tin isotopes, down to \elem{Sn}{96}, see Refs.~\cite{Ireland2024,Ireland2026}. The corresponding two-neutron separation energy in \elem{Sn}{98} across the $N=50$ shell closure is $S_{2n} = 33 \pm 2 \,  \MeV$.
Also, \ai{} nuclear lattice EFT (NLEFT) calculations of \elem{Sn}{99-102} have been recently performed, serving as an independent validation of the capacity of chiral EFT interactions to accurately predict heavy-mass nuclei~\cite{hild25NLEFTSn}.
Theory uncertainties for two-neutron separation energies are assessed by combining the spread between the BCCSD and BCCSD[T] and varying the underlying HO frequency. The combined uncertainty is between $1-2 \, \MeV$ in all nuclei. 

\begin{figure}[t!]
    \centering    \includegraphics[height=6.4cm]{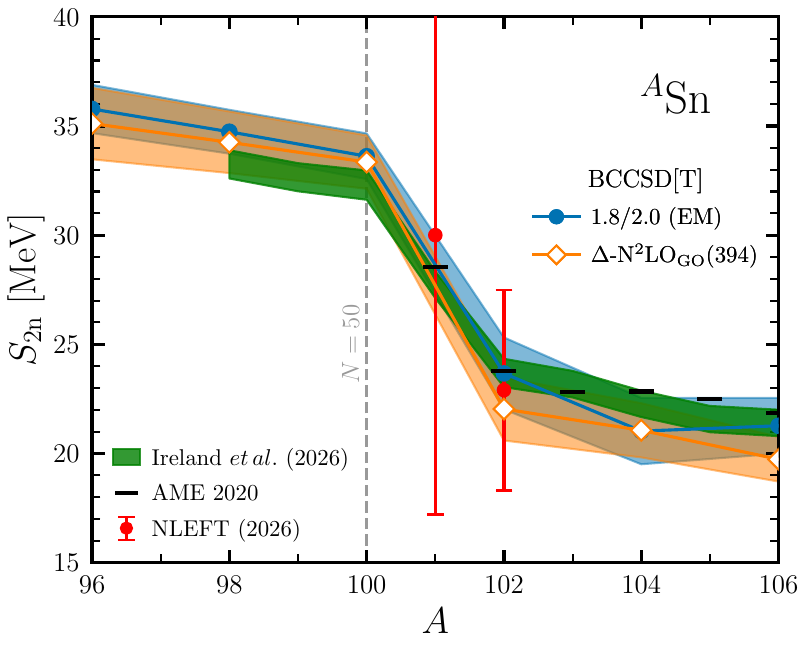}
    \caption{Two-neutron separation energies around the $N=50$ shell closure (\elem{Sn}{100}). BCCSD[T] calculations are shown for \magicint{} (blue) and \deltago{} (orange). 
    NLEFT calculations are taken from Ref.~\cite{hild25NLEFTSn}.
    }
    \label{fig:tin_n50}
\end{figure}

Experimental and theoretical $S_{2n}$ are displayed in Fig.~\ref{fig:tin_n50} over the range \elem{Sn}{96-106}. For both employed Hamiltonians, BCCSD[T] predictions correctly capture the experimental trend for \elem{Sn}{>100} and are consistent with the available NLEFT value in \elem{Sn}{102}. While results for the \magicint{} Hamiltonian are on par with experimental values, the \deltago{} Hamiltonian tends to underpredict two-neutron separation energies by a few MeV, thus exaggerating the $N=50$ shell closure. Across the $N=50$ shell closure, BCCSD[T] predictions are in excellent agreement with extrapolated values from Ref.~\cite{Ireland2026} leading to $S_{2n}(48,50) \approx 33-34 \, \MeV$.

\paragraph*{Chiral interactions at the neutron-rich extremes.--}
\label{s:results}
We finally turn to the prediction of the neutron dripline of even-even tin isotopes that is expected far beyond the heaviest measured nucleus \elem{Sn}{138}. Empirical models such as the liquid drop model and the Duflo-Zuker mass model position the neutron dripline of around $A \approx 150 \pm 5$~\cite{Storbacke2024}. This is in conflict with EDF calculations predicting the dripline to be located around $A=174-178$, \ie{}, close to the $N=126$ shell closure~\cite{Erle12Nature,Neufcourt:2020nme}. While the precise position of the dripline is sensitive to details of the EDF parameterization, the absence of a constructive framework to build EDFs forbids the assessment of associated systematic errors. The large dimension of shell-model spaces makes VS-IMSRG calculations well beyond $N=82$ challenging. Similarly, the high computational cost of NLEFT calculations make them impractical for survey calculations. So far, the BCC framework is the only \ai{} approach in use to target tin isotopes towards the dripline~\cite{Tichai2024bcc,Demol2026bcc}.

\begin{figure}[t!]
    \centering
    \includegraphics[height=6.4cm]{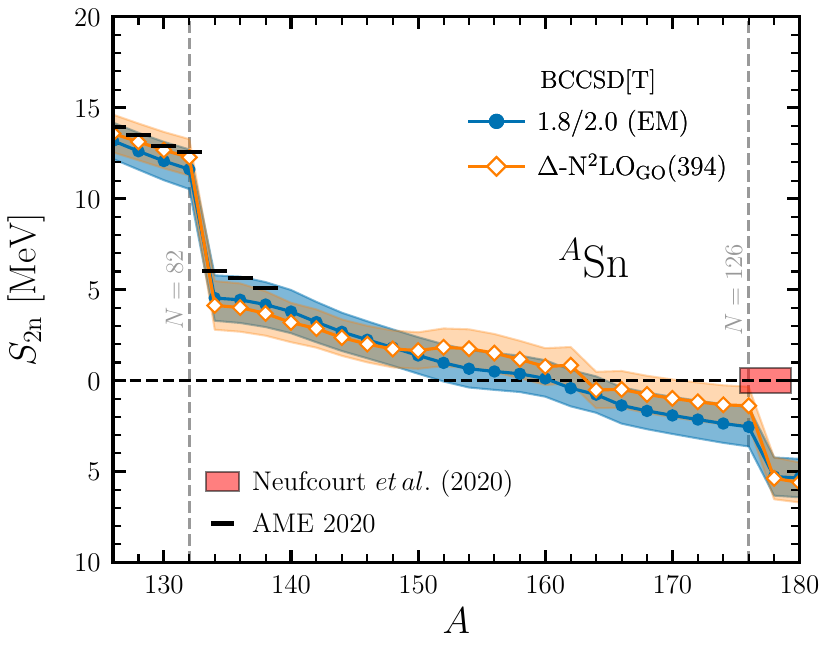}
    \caption{Two-neutron separation energies calculated in BCCSD[T] beyond the $N=82$ shell closure (\elem{Sn}{132}). Calculations details are similar to Fig.~2.}
    \label{fig:tin_dripline}
\end{figure}

The \ai{} predictions based on BCCSD[T] calculations can be inferred from Fig.~\ref{fig:tin_dripline}. Within estimated many-body uncertainties, the neutron drip-line is predicted to be located in the interval $A\approx150-170$ for the \magicint{} Hamiltonian and  $A\approx 160-176$ for the \deltago{} Hamiltonian~\footnote{For the the nucleus \elem{Sn}{162}, we employ for the triples correction the average value obtained from calculations at $\hbar \omega=10,12$. The BCCSD solution yields strong reference sensitivity (large $||\cluster{1}||$ norms). While effects are invisible for the total ground-state energy differential quantities like $S_{2n}$ are more strongly effected.}. The location of the drip-line happens to be {\it fine-tuned} due to the flatness of two-neutron separation energies over a large interval of neutron numbers, which eventually reflects into a large uncertainty. On the many-body side, resolving this fine-tuning will mainly require reducing the uncertainty associated with both normal-ordering approximations of the three-nucleon interaction and uncertainties from basis truncations. The latter manifests in part due to the proximity of the particle continuum that needs to be fully accounted for~\cite{Hu2020gamow}. 
While both state-of-the-art $\chi$EFT Hamiltonians provide rather consistent predictions, there probably exists a significant systematic error associated with the employed nuclear interactions given that experimental two-neutron separation energies immediately after \elem{Sn}{132} are presently underestimated by about $3 \, \MeV$. Propagating (naively) this exaggerated $N=82$ magic character to more neutron-rich isotopes may support a shift of the predicted drip-line by about ten units, thus calling for a proper assessment of interaction uncertainties associated with $\chi$EFT Hamiltonians.

\paragraph*{Conclusions.--}
\label{s:conclusion}
In this work, \ai{} predictions of nuclear masses across the entire tin isotopic chain are performed by advancing the recently proposed \ai{} Bogoliubov coupled-cluster approach to higher precision. This reduces the many-body uncertainty by about a factor of ten, leading to an estimated error on ground-state energies below $1 \, \%$, considerably improving previous predictions.
The two sets of chiral inter-nucleon interactions reproduce the $N=50$ shell closure in agreement with extrapolations from recent high-precision mass measurements. The  location of the neutron drip line is shown to be highly fine-tuned and thus currently uncertain by about 30 neutron numbers. 
The strong sensitivity of $r$-process reaction networks to separation energies of neutron-rich tin isotopes calls for accurate first-principle calculations~\cite{Kuske2025}. The present analysis demonstrates both the significant progress made towards achieving this goal thanks to theory developments and also highlights the immense challenge that remains to be overcome.

\section*{Data availability}

The data presented in this work are openly available on \textsc{Zenodo}~\cite{Vernik26ZenodoTriples}. 

\section*{Acknowledgments}
The authors thank C.~Ireland for sharing the data from Ref.~\cite{Ireland2026} as well as B.~Bally, T.~Papenbrock, A.~Schwenk and L.~Zurek for useful discussions.
This work was supported in part by the LOEWE Top Professorship LOEWE/4a/519/05.00.002(0014)98 by the State of Hesse. This work was supported in part by the European Research Council (ERC) under the European Union's Horizon Europe research and innovation programme (Grant Agreement No.~101162059), by Fonds de la Recherche Scientiﬁque (F.R.S.-FNRS, Belgium) under the MIS Project nr. 40028446 and by Research Foundation Flanders (FWO, Belgium, grant 11G5123N).
The authors gratefully acknowledge the Gauss Centre for Supercomputing e.V. (www.gauss-centre.eu) for funding this project by providing computing time through the John von Neumann Institute for Computing (NIC) on the GCS Supercomputer JUWELS at J\"ulich Supercomputing Centre (JSC).

\section{End matter}
\subsection{Bogoliubov coupled-cluster theory }

In BCC theory~\cite{Sign14BogCC}, the fully correlated state is related to the Bogoliubov reference state via the action of the wave operator
\begin{align}
    |\Psi\rangle \equiv e^\mathcal{T}|\Phi\rangle\,,
\end{align}
where the connected cluster operator $\mathcal{T} \equiv \sum_{n} \mathcal{T}_n$ involves components of different excitation rank $n$~\cite{Sign14BogCC}
\begin{equation}\label{e:Clusteramp}
    \mathcal{T}_n \equiv \frac{1}{(2n)!}\sum_{k_i\dots k_{2n}}t_{k_1\dots k_{2n}}\beta^\dagger_{k_1}\dots\beta^\dagger_{k_{2n}}\, .
\end{equation}

As the Bogoliubov reference state breaks particle-number conservation, the Hamiltonian is replaced by the grand-canonical potential $\Omega \equiv H - \lambda_N N - \lambda_Z Z\,$ where the neutron and proton chemical potentials $\lambda_N$ and $\lambda_Z$ serve as Lagrange multipliers to constrain the average number of neutrons and protons to match the physical values of the target system. 

The BCC approximation with singles and doubles (BCCSD) is defined as the truncation of the cluster operator by
\begin{align}
    \mathcal{T}_\text{BCCSD} \equiv \mathcal{T}_1 + \mathcal{T}_2 \, .
\end{align} 
Introducing  the \emph{similarity-transformed grand potential}
$\bar{\Omega}_N \equiv e^{-\mathcal{T}}\Omega_N e^{\mathcal{T}}$, with $\Omega_N \equiv \Omega - \langle \Phi | \Omega | \Phi \rangle$, the cluster amplitudes are obtained by numerically solving the amplitude equations
\begin{subequations}
\label{eq:BCCSDeq}
    \begin{align}
     S^{\text{BCCSD}}_{\alpha\beta}(\bar{\Omega}_N) &\equiv   \bra{\Phi^{\alpha\beta}}\bar{\Omega}_N\ket{\Phi} = 0 \, ,\\
    D^{\text{BCCSD}}_{\alpha\beta\gamma\delta}(\bar{\Omega}_N) &\equiv    \bra{\Phi^{\alpha\beta\gamma\delta}}\bar{\Omega}_N\ket{\Phi} = 0 \, ,
    \end{align}
\end{subequations}
where reference elementary excitations denote quasi-particle excitations of the Bogoliubov state $|\Phi^{\alpha\beta\dots}\rangle \equiv \beta^{\dagger}_{\alpha}\beta^{\dagger}_{\beta}\dots|\Phi\rangle$. Equations~\eqref{eq:BCCSDeq} determine $\mathcal{T}_1$ and $\mathcal{T}_2$ by decoupling the reference state from its elementary quasi-particle excitations and can be solved at a mild computational cost of $\bigO(N^6)$, where $N$ denotes the size of the computational basis. 

Corrections to observables beyond the Bogoliubov reference state can then be computed, \eg{} the correlation energy is evaluated using the $\Lambda$ approach~\cite{Shav09MBmethod} according to
\begin{align}\label{eq:lambdaE}
    E_\text{corr.} = \bra\Phi(1+\Lambda) \bar{\Omega}_N \ket{\Phi}_{\text{C}}\,.
\end{align}
The de-excitation operator $\Lambda$ is defined as $\Lambda \equiv \sum_{n} \Lambda_n$, with
\begin{align}
    \Lambda_n \equiv \frac{1}{(2n)!}\sum_{k_i\dots k_{2n}}\lambda_{k_1\dots k_{2n}}\beta_{k_{2n}}\dots\beta_{k_1}\, .
\end{align}
Throughout this work, the de-excitation operator is truncated at the doubles level ($\Lambda \approx\Lambda_1 + \Lambda_2$) and approximated as $\Lambda \approx\mathcal{T}^{\dagger}$. Equation \eqref{eq:lambdaE} then becomes 
\begin{align}
    E_\text{corr.} &= \langle \Phi | \bar{\Omega}_N | \Phi \rangle_{\text{C}} + \frac{1}{2!} \sum_{\alpha \beta}t^{*}_{\alpha \beta} \langle \Phi^{\alpha \beta} | \bar{\Omega}_N | \Phi \rangle_{\text{C}}\nonumber\\
    &\quad+ \frac{1}{4!} \sum_{\alpha \beta \gamma \delta}t^{*}_{\alpha \beta\gamma \delta} \langle \Phi^{\alpha \beta\gamma \delta} | \bar{\Omega}_N | \Phi \rangle_{\text{C}} \, \label{e:LambdaCorrelationExplicit} .
\end{align}
Analogous expressions are available for other observables.

\subsection{BCC with singles, doubles and triples}

Higher accuracy in $|\Psi\rangle$ is obtained by including higher-body operators in $\mathcal{T}$; BCC with singles, doubles and triples (BCCSDT) is defined by extending the BCCSD ansatz to
\begin{equation}
    \mathcal{T}_\text{BCCSDT} \equiv \mathcal{T}_\text{BCCSD} + \mathcal{T}_3 \, ,
\end{equation}
where 
\begin{align}
    \mathcal{T}_3 = \frac{1}{6!} \sum_{\alpha \beta \gamma \delta \epsilon \varphi} t_{\alpha \beta \gamma \delta \epsilon \varphi} \,
    \beta^\dagger_{\alpha} \beta^\dagger_{\beta} \beta^\dagger_{\gamma} \beta^\dagger_{\delta} \beta^\dagger_{\epsilon} \beta^\dagger_{\varphi} \, 
\end{align}
encodes the irreducible three-body correlations. The presence of \cluster{3} components results in new terms appearing in Eqs. \eqref{eq:BCCSDeq}
\begin{align}
 S^{\text{BCCSDT}}_{\alpha\beta}(\bar{\Omega}_N)   \equiv \ &  S^{\text{BCCSD}}_{\alpha\beta}(\bar{\Omega}_N)   \nonumber \\
 &+ \langle \Phi^{\alpha \beta} | \Omega_N\mathcal{T}_3 | \Phi \rangle _{\text{C}} \nonumber \\
 = \ & 0\label{eq:singlesintriples}\, ,\\
  D^{\text{BCCSDT}}_{\alpha\beta\gamma\delta}(\bar{\Omega}_N)       \equiv \ & D^{\text{BCCSD}}_{\alpha\beta\gamma\delta}(\bar{\Omega}_N) \nonumber\\ 
        &+ \langle \Phi^{\alpha \beta \gamma \delta} | \Omega_N\mathcal{T}_3 | \Phi\rangle_\text{C} \nonumber \\
        &+ \langle \Phi^{\alpha \beta \gamma \delta} | \Omega_N\mathcal{T}_1\mathcal{T}_3 | \Phi\rangle_\text{C} \nonumber \\
        = \ & 0\, ,\label{eq:doublesintriples}
\end{align}
along with the appearance of the six-quasi-particle amplitude equation to be solved along with the previous two
\begin{align}
\label{eq:tripleseq}
 T^{\text{BCCSDT}}_{\alpha\beta\gamma\delta\epsilon\zeta}(\bar{\Omega}_N)   &\equiv \bra{\Phi^{\alpha\beta\gamma\delta\epsilon\zeta}}\bar{\Omega}_N\ket{\Phi}_\text{C} \nonumber \\
 &= 0\, .
\end{align}
This equation scales as $\bigO(N^8)$, making a full BCCSDT solution computationally prohibitive. Based on perturbative arguments, it is expected that not all terms are equally important and that accurate approximate triples methods with reduced  $\bigO(N^7)$ cost can be designed and implemented.

\subsection{Approximate triples methods}

\begin{table}[t]
\centering
\begin{tabular*}{\columnwidth}{@{\hspace{.75em}}c@{\hspace{.75em}}|@{\hspace{.75em}}c@{\hspace{.75em}}|@{\hspace{.75em}}c@{\hspace{.75em}}|@{\hspace{.75em}}c@{\hspace{.75em}}}
\hline\hline
\eqref{eq:tripleseq} & BCCSDT-3 & BCCSDT-2 & BCCSDT-1 \\
\hline
$e^{\mathcal{T}_1+\mathcal{T}_2+\mathcal{T}_3}\approx$
& $e^{\mathcal{T}_1+\mathcal{T}_2}$
& $e^{\mathcal{T}_2}$
& $\mathcal{T}_2$ \\
\hline\hline
\eqref{eq:doublesintriples}
& \multicolumn{3}{c}{BCCSDT-a} \\
\hline
$e^{\mathcal{T}_1+\mathcal{T}_2+\mathcal{T}_3}\approx$
& \multicolumn{3}{c}{$e^{\mathcal{T}_1+\mathcal{T}_2}+\mathcal{T}_3$} \\
\hline\hline
\end{tabular*}
\caption{Iterative triples methods along with the defining approximation to the wave operator.}
\label{tab:triplesapprox}
\end{table}

Approximate triples methods reduce the numerical cost of Eq. \eqref{eq:tripleseq} while also potentially do so for \eqref{eq:doublesintriples} at the same time. Comprehensive reviews of triples corrections in standard CC theory from a quantum-chemistry perspective can be found in Refs.~\cite{Shav09MBmethod, Urban85CC, RAGHAVACHARI1989479, TaubeLambdaCC, KucharskiNonIterativeT} and from a nuclear physics perspective in Refs.~\cite{Demol2024Thesis, Vernik2024Master}. Here, we formulate the approximate inclusion of triples corrections in the context of Bogoliubov CC for the first time. There are essentially two classes of methods to do so. 

A first option consists of approximating the exponential ansatz in Eq.~\eqref{eq:tripleseq} after having transformed the latter into a fixed-point equation for $\mathcal{T}_3$. This corresponds to using one of the T-3, T-2 or T-1 simplifications listed in Tab. \ref{tab:triplesapprox}. This can be combined with the T-a approximation also listed in Tab. \ref{tab:triplesapprox} to solve for $\mathcal{T}_2$ via \eqref{eq:doublesintriples}. This first class of approximations still requires repeated evaluations of $\bigO(N^7)$ equations and several copies of $\mathcal{T}_3$ to be stored.

Alternatively, a non-iterative evaluation of triples contributions can be performed: one first solves BCCSD equations and then inputs the resulting \cluster{1}, \cluster{2} amplitudes into Eq. \eqref{eq:tripleseq} to obtain $\mathcal{T}_3$ \textit{a posteriori}. Using the T-1a approximation one obtains the simplest non-iterative approach to $\mathcal{T}_3$ given by
\begin{equation}
    t_{\alpha\beta\gamma\delta\epsilon\zeta} = -\mathcal{A} \sum_{k_{1}} \frac{\Omega^{31}_{\delta\epsilon\zeta k_{1}}\,t_{\alpha\beta\gamma k_{1}}}{\Delta E_{\alpha\beta\gamma\delta\epsilon\zeta}}
    \label{e:triplesBCCSDT1a}\, ,
\end{equation}
where $\mathcal{A}$ denotes the anti-symmetrizer ensuring fermionic symmetry and where  $\Omega^{31}$ is a specific component of $\Omega$ expressed in the quasi-particle basis~\cite{Sign14BogCC}. The energy denominator involves the sum of quasi-particle energies 
\begin{align}
    \Delta E_{\alpha\beta\gamma\delta\epsilon\varphi}\equiv E_\alpha + E_\beta + E_\gamma + E_\delta + E_\epsilon + E_\varphi \, .
\end{align}
The corresponding BCCSD[T] energy correction is then obtained by using the $\Lambda$ approach (see Eq. \eqref{e:LambdaCorrelationExplicit}) and by keeping only what was shown to be the dominant energy contribution due to triples,
\begin{align}
    E_\text{t}
    &\equiv -\frac{1}{6!}\sum_{\alpha\beta\gamma\epsilon\zeta\eta}\,t^{*}_{\alpha\beta\gamma\epsilon\zeta\eta}\,\Delta E_{\alpha\beta\gamma\epsilon\zeta\eta}\,t_{\alpha\beta\gamma\epsilon\zeta\eta}\label{e:e_t{T}2} \, ,
\end{align}
which involves $\bigO(N^6)$ operations whose cost is thus similar to solving BCCSD equations.
However, the computationally most expensive part is the formation of the triples amplitudes themselves (see Eq.~\eqref{e:triplesBCCSDT1a}) that involves $\bigO(N^7)$ operations and hence exceeds the cost of solving the BCCSD amplitude equations. 
Due to its non-iterative character, the proposed BCCSD[T] framework does not require the storage of multiple copies of the $\cluster{3}$ amplitudes. 
The memory consumption of BCCSD[T] calculations is thus still driven by the \cluster{2} amplitudes and the storage of the corresponding DIIS history that is used to accelerate convergence~\cite{Pulay1980diis}. 

One can relax the approximations made in BCCSD[T] by considering all the terms in the $\Lambda$ evaluation of the triples correction to the energy while using triples given in Eq. \eqref{e:triplesBCCSDT1a}. This yields two additional sub-leading contributions
\begin{subequations}\label{e:(T)EnergyCorrections}
    \begin{align}
        E_{\text{st}}  &= \frac{1}{2!}\frac{1}{4!} \sum_{\alpha\beta}t^*_{\alpha\beta}\sum_{\gamma\delta\epsilon\zeta}\Omega_{\gamma\delta\epsilon\zeta}^{04}\, t_{\alpha\beta\gamma\delta\epsilon\zeta}\, ,\label{e:E_st}\\
        E_{\text{dt}}  &= \frac{1}{4!}\frac{1}{2!}\sum_{\alpha\beta\gamma\delta}t^*_{\alpha\beta\gamma\delta}\sum_{\epsilon\zeta}\Omega_{\epsilon\zeta}^{02}\, t_{\alpha\beta\gamma\delta\epsilon\zeta}\label{e:E_dt}\, ,
    \end{align}
\end{subequations}
defining the so-called BCCSD(T) approximation that is complete up to fourth order in Bogoliubov many-body perturbation theory (BMBPT). The Bogoliubov framework relies on breaking particle-number (PN) symmetry that ultimately needs to be restored to ensure that the system of interest is made from the correct integer number of neutrons and protons. It turns out that BCCSD(T) introduces a PN shift that cannot be consistently corrected for in a non-iterative fashion. To overcome this issue while retaining BMBPT completeness up to fourth order, the approximation on $\mathcal{T}_3$ is relaxed from T-1a to T-2a, thus leading to the BCCSD\{T\} approximation. The corresponding expression of triples amplitudes is
\begin{equation}
    t_{\alpha\beta\gamma\delta\epsilon\zeta} = -\mathcal{A} \sum_{k_{1}} \frac{\chi^{31}_{\delta\epsilon\zeta k_{1}}\,t_{\alpha\beta\gamma k_{1}}}{\Delta E_{\alpha\beta\gamma\delta\epsilon\zeta}}
    \label{e:triplesBCCSDT2a}\, ,
\end{equation}
where the intermediate $\chi^{31}$ is given by
 \begin{align}\label{e:ChiDef}
        \chi^{31}_{\delta\epsilon\zeta k_1}(\Omega) &= \Omega^{31}_{\delta\epsilon\zeta k_{1}}+ \frac{1}{2!}\sum_{k_2}\Omega^{02}_{k_1 k_2}\,t_{k_2 \delta\epsilon\zeta}\nonumber\\
        &\quad+ \frac{1}{2!}P(\delta/\epsilon\zeta)\sum_{k_2 k_3}\Omega^{13}_{\delta k_1 k_2 k_3}\,t_{k_2k_3 \epsilon\zeta}\, .
    \end{align}
The expressions for the correlation energy remain formally unchanged from BCCSD(T). The overview of the non-iterative triples methods is given in Tab. \ref{tab:noniterativemethos}.
\begin{table}[t!]
    \centering
    \begin{tabular*}{\columnwidth}{@{\hspace{.30em}}c@{\hspace{.30em}}|@{\hspace{.30em}}c@{\hspace{.30em}}|@{\hspace{.30em}}c@{\hspace{.30em}}|@{\hspace{.30em}}c@{\hspace{.30em}}}
    \hline\hline
         & BCCSD\{T\} & BCCSD(T) & BCCSD[T] \\
         \hline
         Def. of triples & \eqref{e:triplesBCCSDT2a} &  \eqref{e:triplesBCCSDT1a} &\eqref{e:triplesBCCSDT1a}\\
         Exp. for $E$ & \eqref{e:e_t{T}2} $+$ \eqref{e:(T)EnergyCorrections} & \eqref{e:e_t{T}2} $+$ \eqref{e:(T)EnergyCorrections} & \eqref{e:e_t{T}2}\\
         BMBPT(4) complete & Yes & Yes & No\\
         PN shift & Yes & Yes & No\\
         PN constrainable & Yes & No & N/A\\
         \hline\hline
    \end{tabular*}
    \caption{Non-iterative triples methods along with the corresponding definition of six-quasi-particle amplitudes, expression for the correlation energy, and their BMBPT and PN properties.}
    \label{tab:noniterativemethos}
\end{table}

\bibliographystyle{apsrev4-2}
\bibliography{strongint}

\end{document}